\documentclass[
 amsmath,amssymb,
twocolumn
]{revtex4-2}

\usepackage{graphicx}
\usepackage{bm}

\begin{document}


\title{\textbf{Sub-Hz Stability and Correlation in Pair-Generated Primary Kerr Comb Tones}}

\author{Konstantin Khrizman}
\email{Contact author: konstantin.khrizman@mail.huji.ac.il}
\author{Andrei Diakonov}%
\author{Liron Stern}
 
\affiliation{Applied Physics Institute, Hebrew University of Jerusalem, Jerusalem, Israel}

\begin{abstract}

Kerr microcombs provide a compact route to broadband optical frequency grids, yet the primary comb states formed at the onset of Kerr-comb generation have received little attention as metrological objects. Here we characterize the coherence and frequency stability of pair-generated primary-comb tones in a silicon nitride microresonator using synchronized multi-channel frequency counting referenced to a hydrogen-maser-stabilized difference-frequency comb, enabling direct measurement of temporal fluctuations and correlations among the pump, signal, and idler tones. We show that the generated tones are strongly constrained by parametric energy conservation: under weakly locked conditions with MHz-level frequency excursions, the residual deviation from $2f_p=f_s+f_i$ remains sub-hertz in the mean, and the signal-idler regression deviates from the ideal $-1$ response by only $2.4\times10^{-9}$. When two of the three tones are tightly phase-locked, the energy-conservation residual of the full pump-signal-idler triad, equivalently the deviation of the measured idler from the value inferred from the locked pump and signal, reaches a fractional-instability floor near $6 \times 10^{-16}$ at $\tau\approx100~\mathrm{s}$. This demonstrates metrological-level preservation of the parametric constraint while revealing subtle mode-dependent noise transfer. Together, these results establish primary Kerr tones as a strongly correlated chip-scale parametric frequency triad suitable for demanding precision-frequency applications.

\end{abstract}

\maketitle

\section{Introduction}

The nonlinear process of four-wave mixing (FWM) in high-Q micro-resonators is a fundamental mechanism that initiates 
optical parametric oscillation, driving the growth of signal and idler fields from vacuum fluctuations \cite{kippenberg2004kerr,savchenkov2004low,delhaye2007optical,kippenberg2011microresonator,kippenberg2018dissipative, chembo2013spatiotemporal}. This initial stage of pattern formation gives rise to the primary comb regime \cite{herr2012universal,godey2014stability, Diallo2017PrimaryKerrCombs, pfeifle2015optimally}, a distinct dynamical state characterized by strong 
parametric lines spaced by multiple free spectral ranges $(\mu \times \text{FSR})$. This initial stage is closely related to optical parametric oscillation in a Kerr cavity: above threshold, a pump photon pair is converted into signal and idler photons whose frequencies are set by phase matching and resonator dispersion \cite{kippenberg2004kerr,herr2012universal}. A prominent realization of these dynamics is the Kerr micro-resonator frequency comb, which has become a central platform for chip-scale nonlinear photonics \cite{gaeta2019photonic,pasquazi2018microcombs} 

Despite the extensive focus on soliton microcombs, primary Kerr combs remain under-characterized from a precision-metrology perspective, even though they are expected to exhibit a rigid energy-conservation relation and favorable phase-noise characteristics \cite{liang2015high,huang2015low,li2012low}. While full soliton state often demand complex pumping sequences and broad spectral management, primary combs offer a robust, high-power, reduced-dimensional testbed for parametric physics and a useful state for applications that do not require ultra-broad bandwidths. Yet, unresolved questions remain: to what extent are the measured primary-comb lines mutually correlated? How are measured frequency fluctuations distributed among the different modes?

Understanding how frequency noise is shared among the pump, signal, and idler addresses both fundamental and practical challenges. In an ideal parametric oscillator, the generated signal and idler are constrained by energy conservation, $2f_p=f_s+f_i$, so that their fluctuations remain mutually correlated. In realistic microresonators, however, this constraint may be degraded by pump technical noise, thermorefractive fluctuations \cite{huang2019thermorefractive,kondratiev2018thermorefractive}, Raman scattering, mode coupling, and the proximity of primary states to unstable or chaotic comb dynamics. Quantifying the limits of this mutual coherence is therefore essential for assessing whether primary comb states can act as synchronized, metrologically useful frequency grids, or whether their generated sidebands are ultimately limited by independent noise channels.

The metrological and practical value of primary combs depends on whether the pump, signal, and idler form a genuinely rigid frequency triad. If the generated sidebands remain constrained by energy conservation, stabilizing any two tones determines the third, enabling stability transfer, frequency translation, and common-mode noise rejection across large optical frequency separations \cite{coillet2014robustness}. Unlike fully developed comb states, primary combs consist of a small number of well-isolated, high-power tones rather than a dense spectrum of closely spaced lines \cite{pfeifle2015optimally}. This makes them a simplified metrological platform for frequency transfer across tens of nanometers, based on a sparse set of parametrically linked optical frequencies.

In this work, we present a systematic metrological characterization of the synchronized frequency dynamics of a primary Kerr comb generated in a silicon nitride ($\mathrm{Si_3N_4}$) micro-resonator. We use synchronized multi-channel frequency counting, referenced to a hydrogen-maser-stabilized optical difference frequency comb (DFC), to directly measure the temporal frequency fluctuations and correlations of the pump, signal, and idler tones. Separately, we analyze the phase noise of the relevant heterodyne beats to characterize the short-time coherence of the primary tones. Under weakly locked conditions, where the comb is effectively free running, the individual optical frequencies fluctuate at the MHz level, yet the residual deviation from $2f_p=f_s+f_i$ remains bounded at the sub-hertz level. We further implement tight, simultaneous phase locking of the pump and signal tones, allowing the residual correlation to be tested under active stabilization. The stabilized residual reaches a fractional-instability floor near $6 \times 10^{-16}$, while exposing a subtle noise asymmetry between the generated sidebands. Together, these results identify primary Kerr tones as coherent, frequency-rigid optical grids for precision metrology.

\section{Experimental System}

The experimental layout is designed to measure the three optical frequencies that define the primary-comb triad: the pump and the generated signal and idler sidebands. As illustrated in Fig.~\ref{fig:setup}, each tone is optically filtered, heterodyned against a common hydrogen-maser-referenced difference-frequency comb, and recorded by synchronized frequency counters. This common-reference, multi-channel measurement allows the residual $f_s + f_i - 2f_p$ to be reconstructed directly from the time-domain frequency records.

The experimental system is shown in Fig.~\ref{fig:setup_fig}. The primary comb is generated in a high-$Q$ silicon nitride $(\mathrm{Si_3N_4})$ microresonator with a radius of $R\approx22.5~\mu\mathrm{m}$,  with a $\mathrm{SiO_2}$ top and bottom cladding. The device free spectral range is approximately $1.05~\mathrm{THz}$. To access the primary-comb regime, the cavity is pumped with a continuous-wave laser that is tuned from the blue-detuned side into resonance. The detuning is then set such that the primary sidebands are sufficiently strong for detection, while the system remains below the onset of chaotic modulation-instability dynamics~\cite{godey2014stability}. Throughout the paper, we denote the first generated sidebands as the signal and idler, located at mode numbers $+\mu$ and $-\mu$ relative to the pump, respectively.

To track the frequency fluctuations of the pump, signal, and idler simultaneously, the three optical tones are separated using bandpass filters and independently heterodyned with the a Maser-referenced DFC, which provides a stable optical frequency grid across the C band. Each Kerr-comb tone is mixed with the nearest DFC tooth and detected on a balanced photodetector. To avoid detector saturation, an additional bandpass filter is used to restrict the DFC spectrum to the spectral region near the relevant Kerr-comb tone. In this way, we obtain three simultaneous beat notes, one for each of the pump, signal, and idler tones:

\begin{figure}[!t]
  \includegraphics[width=\linewidth]{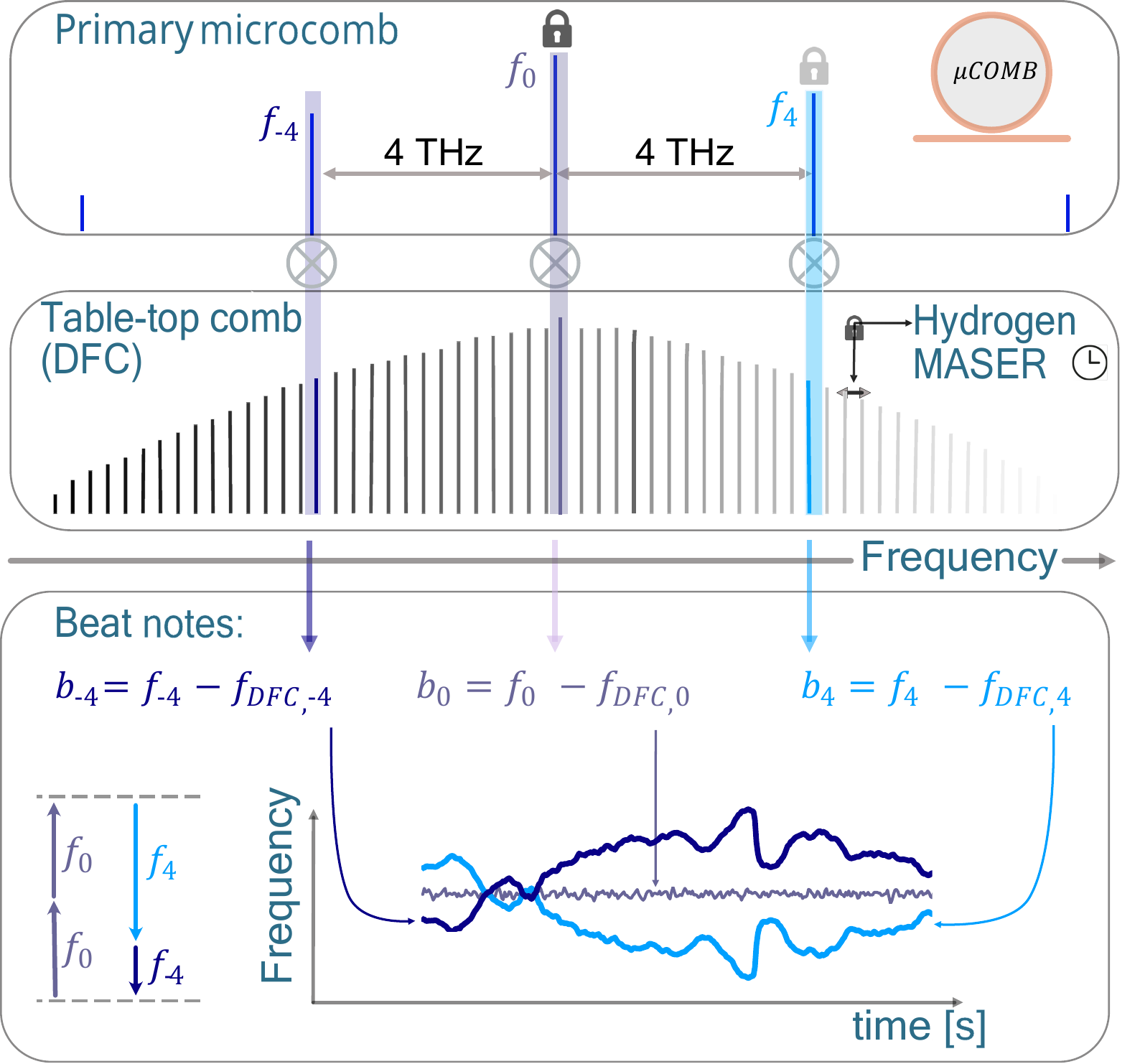}
    \caption{Synchronized tracking of parametric frequency correlations in a primary Kerr comb. A high-$Q$ SiN microresonator is pumped to generate a primary comb with tones at $\mu=0$ and $\mu=\pm4$, corresponding to the pump, signal, and idler. The sidebands are separated from the pump by $4\times\mathrm{FSR}\approx4.20~\mathrm{THz}$. Each tone is filtered and heterodyned with a corresponding tooth of a difference-frequency comb (DFC), whose repetition rate is referenced to a hydrogen maser. The resulting beat notes, $b_{-4}(t)$, $b_0(t)$, and $b_{+4}(t)$, are counted synchronously using a common maser-referenced time base, enabling direct reconstruction of the energy-conservation residual $f_{-4}+f_{+4}-2f_0$.}
  \label{fig:setup}
\end{figure}

\begin{equation}
b_\mu(t)= \left| f_\mu(t)-f^{\mathrm{DFC}}_{k(\mu)} \right|
\end{equation}

\begin{figure*}[!t]
    \centering
    \includegraphics[width=\textwidth]{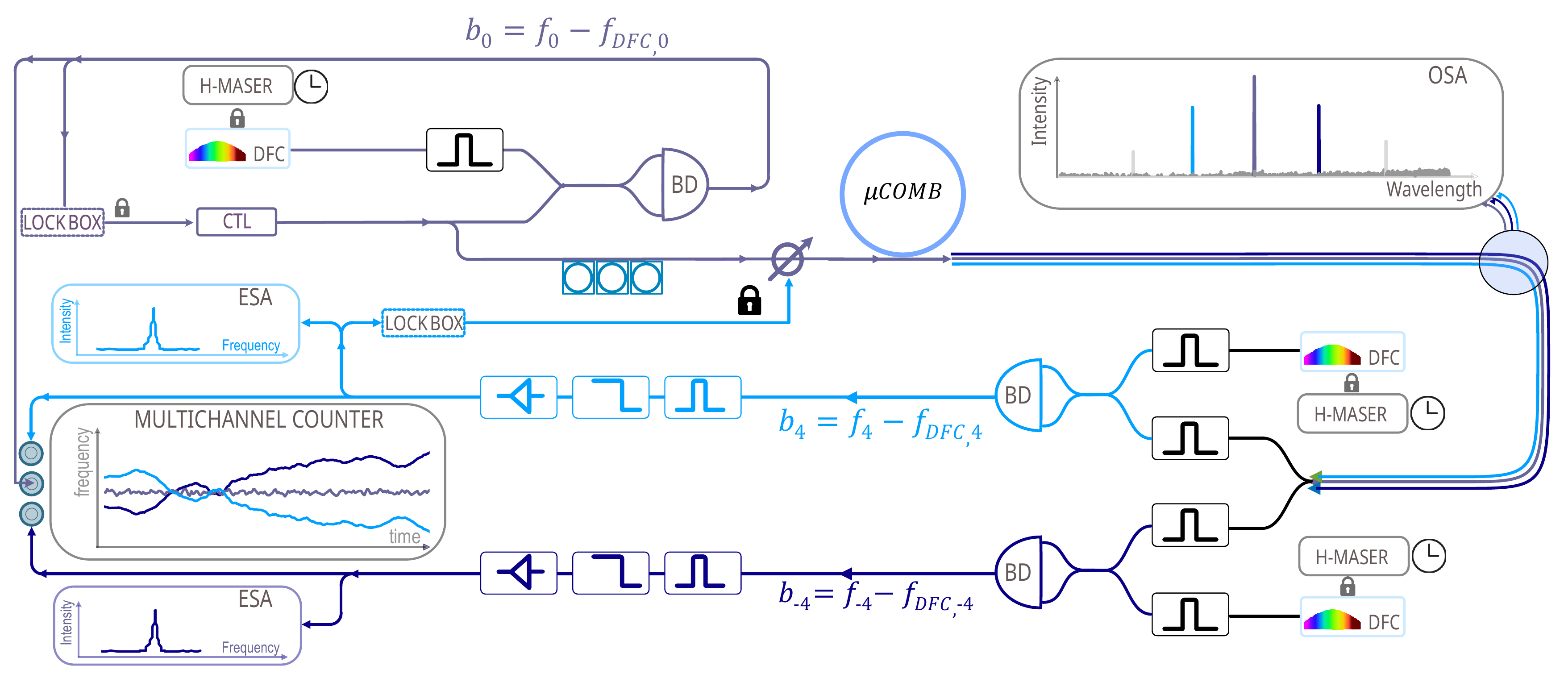}
    \caption{Experimental setup for synchronized tracking of primary-comb frequency correlations. A continuous-wave laser is amplified and coupled into a high-Q SiN microresonator to generate the primary Kerr comb. The pump, signal, and idler tones are separated and independently heterodyned with a tabletop difference-frequency comb (DFC), whose repetition rate is referenced to a hydrogen maser. The resulting beat notes, $b_0=f_0-f_{\mathrm{DFC},0}$, $b_{+4}=f_{+4}-f_{\mathrm{DFC},+4}$, and $b_{-4}=f_{-4}-f_{\mathrm{DFC},-4}$, are detected on balanced photodetectors and recorded using synchronized multi-channel frequency counters. Two phase-locking loops are used to stabilize the pump and signal tones to the common DFC reference, enabling measurement of the residual frequency fluctuations of the idler and reconstruction of the energy-conservation residual $2f_0-f_{-4}-f_{+4}$.}
    \label{fig:setup_fig}
\end{figure*}

where $f_\mu(t)$ is the optical frequency of the Kerr-comb tone at mode index $\mu$, and $f^{\mathrm{DFC}}_{k(\mu)}$ is the frequency of the DFC tooth used for the corresponding heterodyne measurement.

The three beat notes, $b_{-4}(t)$, $b_0(t)$, and $b_{+4}(t)$, are recorded synchronously using a common dead-time-free $\Lambda$-type multi-channel counter referenced to the same hydrogen-maser time base. This ensures time-aligned sampling across all channels. To improve the beat-note quality, each RF signal is amplified and bandpass filtered using an RF bandpass filter with a bandwidth of approximately $500~\mathrm{kHz}$, which increases the effective signal-to-noise ratio of the detected beats.

\begin{figure*}[!t]
    \centering
    \includegraphics[width=\textwidth]{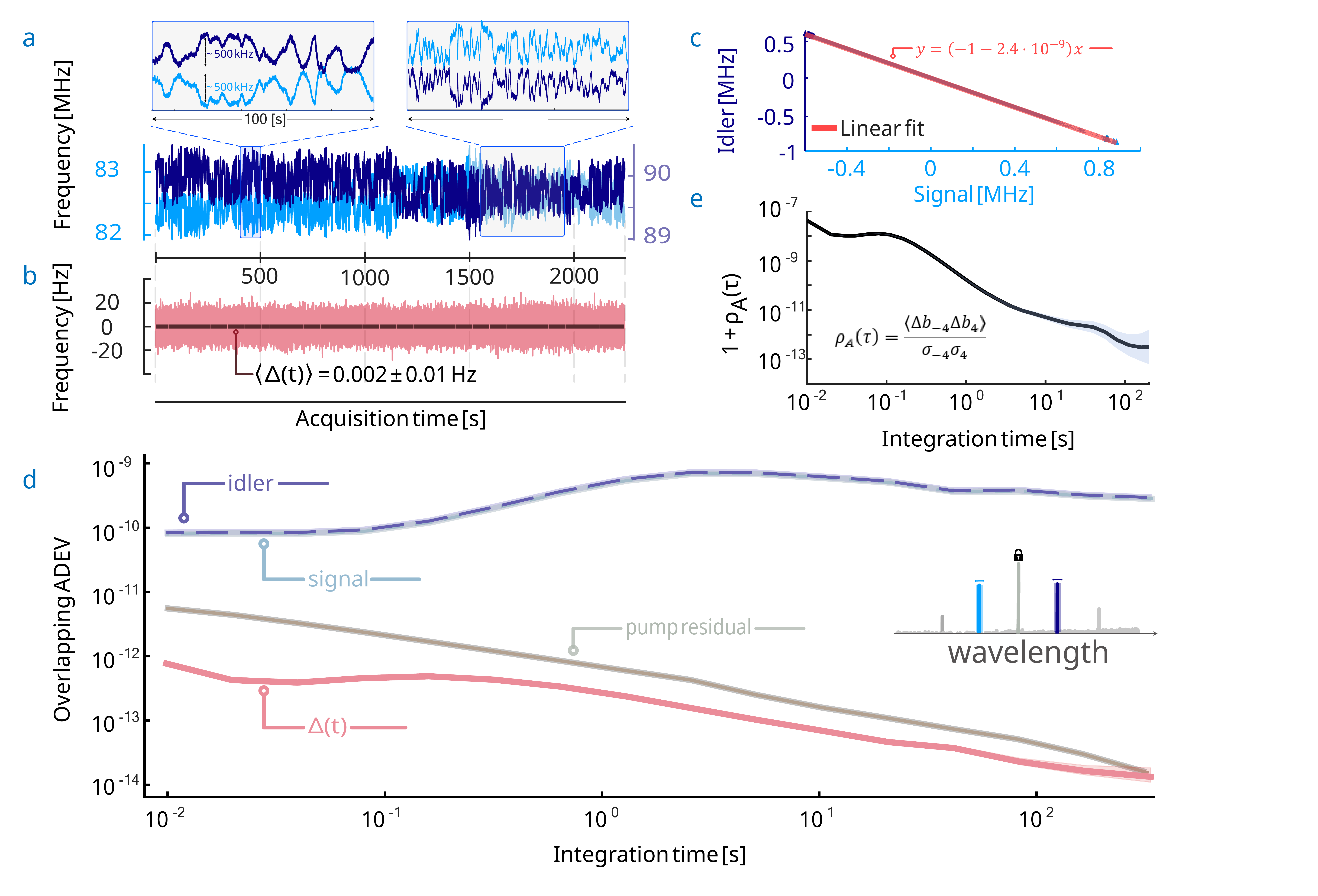}
    \caption{\textbf{Synchronized multi-channel frequency characterization of a primary Kerr comb.}
\textbf{(a)} Time-domain frequency records of the signal, $\mu=+4$, and idler, $\mu=-4$, beat notes relative to the DFC, acquired with a gate time of $10~\mathrm{ms}$. The inset panels show representative temporal snapshots, highlighting correlated fluctuations of the generated sidebands.
\textbf{(b)} Energy-conservation residual, defined as $\Delta(t)=b_{+4}(t)+b_{-4}(t)-2b_0(t)$, showing a sub-hertz mean residual and strong rejection of common frequency noise.
\textbf{(c)} Linear regression of the idler beat frequency versus the signal beat frequency, showing the expected anti-correlated response of the generated sidebands.
\textbf{(d)} Overlapping Allan deviation (OADEV) of the normalized pump, signal, and idler beat-note frequencies, together with the normalized energy-conservation residual $\Delta(t)$.
\textbf{(e)} Allan-correlation parameter plotted as $1+\rho_A(\tau)$, where $\rho_A(\tau)\rightarrow -1$ corresponds to near-perfect anti-correlation between the signal and idler frequency fluctuations at long integration times.}
    \label{fig3}
\end{figure*}

For the measurement shown in Fig.~\ref{fig3}, we use a weak-locking configuration. The pump beat note is locked to the DFC using a broadband phase/frequency detector and a fast PID controller. The signal sideband is weakly stabilized using an RF delay-line discriminator: the beat note is split into two paths with different coaxial delays and recombined on a phase detector \cite{schunemann1999simple}. The resulting error signal is applied to the pump power through a variable optical attenuator (VOA), providing only a weak constraint on the primary-comb spacing and on the generated sideband frequencies. This configuration enables direct observation of the correlated signal-idler dynamics under effectively free-running primary-comb conditions.

In Fig.~\ref{fig4}, we then replace this weak sideband stabilization with a commercial fast offset-locking servos, which tightly phase-lock the pump and signal beats to the common DFC reference. This second configuration allows the energy-conservation residual to be tested under tight active stabilization.

\section{Results}
We next present the results of the synchronized multi-channel frequency measurement. A primary comb is generated by pumping the microresonator with continuous-wave light at $1551~\mathrm{nm}$. The first generated sidebands appear at $\sim1519~\mathrm{nm}$ and $\sim1584~\mathrm{nm}$, corresponding to the signal $(\mu=+4)$ and idler $(\mu=-4)$, respectively. Figure~\ref{fig3}a shows a representative synchronized measurement of the pump, signal, and idler beat notes relative to the DFC, acquired over $40~\mathrm{min}$ with a gate time of $10~\mathrm{ms}$. In this measurement, the pump beat note is phase-locked to the DFC, while the $1519~\mathrm{nm}$ signal beat note is weakly stabilized using a delay-line discriminator, whose error signal is fed back to a variable optical attenuator. As a result, the signal beat note remains centered near $82.5~\mathrm{MHz}$, but still exhibits relatively large frequency excursions on the order of $500~\mathrm{kHz}$. The idler beat note, centered near $90.0~\mathrm{MHz}$, follows these fluctuations through the parametric coupling of the four-wave-mixing process. Representative temporal snapshots are shown in the upper panels of Fig.~\ref{fig3}a, illustrating the correlated motion of the signal and idler frequencies.

To quantify this constraint, we reconstruct the energy-conservation residual, $\Delta(t)=b_{+4}(t)+b_{-4}(t)-2b_0(t)$, as shown in Fig.~\ref{fig3}b. Despite the large excursions of the individual sideband beat notes, $\Delta(t)$ remains tightly clustered around zero, with a mean value of $\langle\Delta(t)\rangle=0.002~\mathrm{Hz}$ over the full acquisition time, consistent with zero at the level of the corresponding Allan deviation, which reaches approximately $0.04~\mathrm{Hz}$ at an integration time of about $300~\mathrm{s}$. This sub-hertz residual shows that the dominant frequency fluctuations are shared by the generated sidebands and are rejected in the energy-conservation combination.  

The correlation is further quantified by the linear regression shown in Fig.~\ref{fig3}c, where we plot $b_{-4}$ versus $b_{+4}$. For perfect energy-conserving transfer with a fixed pump, a $1~\mathrm{MHz}$ increase in the signal beat frequency should be accompanied by a $1~\mathrm{MHz}$ decrease in the idler beat frequency, corresponding to an ideal slope of $-1~\mathrm{MHz}/\mathrm{MHz}$. The fitted slope deviates from this ideal value by only $2.4\times10^{-9}~\mathrm{MHz}/\mathrm{MHz}$. Thus, on average over the full data set, the residual transfer error is approximately $2.4~\mathrm{mHz}$ for every $1~\mathrm{MHz}$ of signal-frequency excursion. This confirms that the large signal and idler excursions are nearly perfectly anti-correlated in their average transfer response, with only a minute deviation from the ideal parametric relation.

\begin{figure*}[!t]
    \centering
    \includegraphics[width=\linewidth]{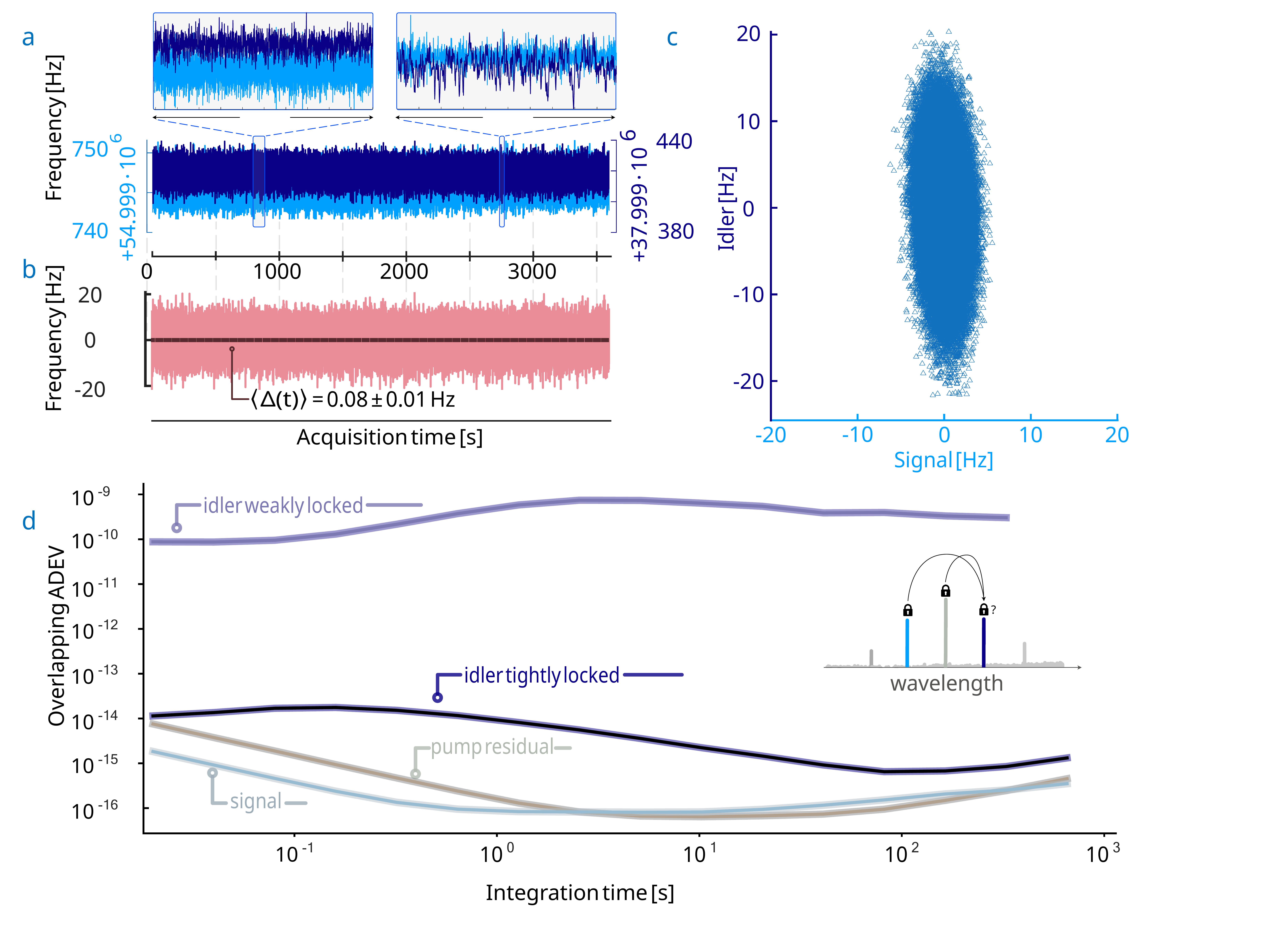}
    \caption{\textbf{High-precision stabilization and noise asymmetry in the primary comb.} 
    \textbf{(a)} 20 ms gate time synchronized measurement with pump and signal frequencies phase-locked to the DFC. The 
    signal frequency fluctuations (cyan) are suppressed to $<10$ Hz, while the idler frequency (navy blue) exhibits 
    residual excursions of approximately 40 Hz.
    \textbf{(b)} Energy conservation residuals $\Delta(t)$ for the high-stability regime.
    \textbf{(c)} Linear regression of the stabilized sidebands
    \textbf{(d)} Overlapping Allan deviation (OADEV) of the normalized pump residual, tightly locked signal and idler compared to the weakly-locked idler.}
    \label{fig4}
\end{figure*}

In Fig.~\ref{fig3}d, we quantify the stability of the measured pump, signal, and idler beat-note frequencies relative to the DFC using the overlapping Allan deviation (OADEV) \cite{dawkins2007considerations}, together with the energy-conservation residual, $\Delta(t)$. Because the optical frequencies were stabilized to the DFC and measured against the same reference, the reported instabilities correspond to residual beat-note fluctuations of the complete DFC-referenced system. These residual fluctuations represent the additive noise introduced by the chip-scale nonlinear comb process, coupling into and out of the chip, and the measurement chain. We note that the pump, signal, and idler beat frequencies were each normalized to their corresponding optical carrier frequencies, while the energy-conservation residual, $\Delta(t)$, was normalized to the sideband-to-sideband frequency separation, approximately $8.40~\mathrm{THz}$. This choice reflects the interpretation of $\Delta(t)$ as the residual mismatch between the two generated sideband spacings. If instead $\Delta(t)$ is interpreted as an absolute deviation from the four-wave-mixing energy-conservation condition, it may also be normalized to an optical carrier frequency. While the weakly locked system exhibits short-term instability limited by thermal fluctuations \cite{huang2019thermorefractive,kondratiev2018thermorefractive} reaching $5\times 10^{-10}$ at $1$~s, the Allan deviations of the idler and signal frequencies show strong correlation, with their curves nearly overlapping. Crucially, the energy conservation residual (red 
curve, $\Delta(t)$) demonstrates the profound parametric rigidity of the primary comb. Despite the pronounced instability of the individual modes, the residual remains orders of magnitude lower and immune to these large-scale excursions. Starting at a fractional instability of roughly $10^{-12}$ at $10^{-2}$~s, it steadily averages down to the $10^{-14}$ level at integration times exceeding 100~s. These values correspond to a fractional instability of roughly $4\times10^{-16}$ when normalized to the optical carrier. We note that, from integration times slightly below $1~\mathrm{s}$ to the longest measured averaging times, the residual decreases approximately as $\tau^{-1/2}$. This suggests that, after rejection of the dominant correlated signal-idler motion, the remaining residual is largely governed by uncorrelated white frequency noise or by the measurement floor.

The behavior is also reflected in Fig. \ref{fig3}e, which plots $ 1 + \rho_A(\tau)$ as a function of integration time.
Here, $\rho_A = \frac{\langle \Delta b_{-4} \Delta b_{4} \rangle}{\sigma_{-4} \sigma_{4}}$ is the overlapping Allan deviation 
correlation coefficient, where  $\sigma_1, \sigma_2 $ denote the individual OADEV of each channel at integration time
$\tau$ \cite{groslambert1981characterization,vernotte2019three}. As $\tau$ increases, the $ 1 + \rho_A(\tau)$ follows a consistent downward trend, reaching a value of 
$10^{-13}$ at $\tau =  10^{2}~s$, indicating near perfect anti-correlation.  

Having established the strong signal-idler correlation under weak locking, we phase-lock the pump and signal beats to the common DFC reference using two commercial fast offset-locking servos, and examine how completely the parametric constraint transfers their stability to the unlocked idler, a question that is relevant both to the fundamental dynamics of primary-comb generation and to frequency-translation applications. Figure~\ref{fig4} presents a one-hour synchronized measurement acquired at a 20~ms gate time under this tight-locking configuration.

The directly servo-controlled signal residual is suppressed below 10~Hz. Because the large $\sim$500~kHz excursions of the weak-locking regime 
are almost entirely common-mode, removing them through the lock leaves the idler constrained only by the parametric coupling, where it settles
onto a residual of $\sim$40~Hz peak-to-peak at the 20~ms gate time. This 40~Hz floor is not produced by the lock. It coincides with the energy-conservation residual band already present in the weakly locked regime (Fig.~\ref{fig3}b), previously masked by the much larger common-mode motion, and that two locked tones pin the third to this pre-existing floor confirms that the parametric constraint transfers stability across the triad. The overlapping Allan deviation shown in Fig.~\ref{fig4}d reflects the same picture: the tightly locked idler averages down with a white-frequency slope, improving by roughly four orders of magnitude relative to the weakly locked case at $\tau \approx 10^{-1}\text{--}10^{2}$~s.

Because both the pump and the signal are phase-locked to the maser-referenced DFC, the energy-conservation constraint maps their stability onto the unlocked idler. Correspondingly, the residual idler beat-note fluctuations average down to a fractional instability of approximately $6\times10^{-16}$ at $\tau\approx100~\mathrm{s}$, referred to the idler carrier frequency.

The transfer is nonetheless imperfect. The idler residual remains about four times larger than the signal ($\sim$40~Hz versus $<$10~Hz), and its OADEV sits roughly an order of magnitude above the signal across the measured range. Consistently, the stabilized signal and idler scatter (Fig.~\ref{fig4}c) shows a loss of instantaneous correlation, collapsing to a weak Pearson coefficient $r = -0.15$, while the mean residual remains sub-hertz, $\langle \Delta(t) \rangle \approx 0.08$~Hz (Fig.~\ref{fig4}b).

From this single-configuration dataset we cannot determine the origin of the
asymmetric idler floor, and several explanations remain consistent with the
data. It may arise from the noise floor of the measurement system itself,
including the RF amplification and detection electronics and the frequency
counter. It may also reflect a differential measurement-path effect, since the signal and idler traverse independent fiber and receiver chains whose noise does not cancel in $\Delta$.
Finally, it may originate from an intracavity mechanism such as the Raman
self-frequency shift or asymmetric thermal pulling. We therefore report the asymmetric idler floor as an observation, while leaving a systematic identification of its origin to a separate study.

\begin{figure}[!b]
    \centering
    \includegraphics[width=1.1\linewidth]{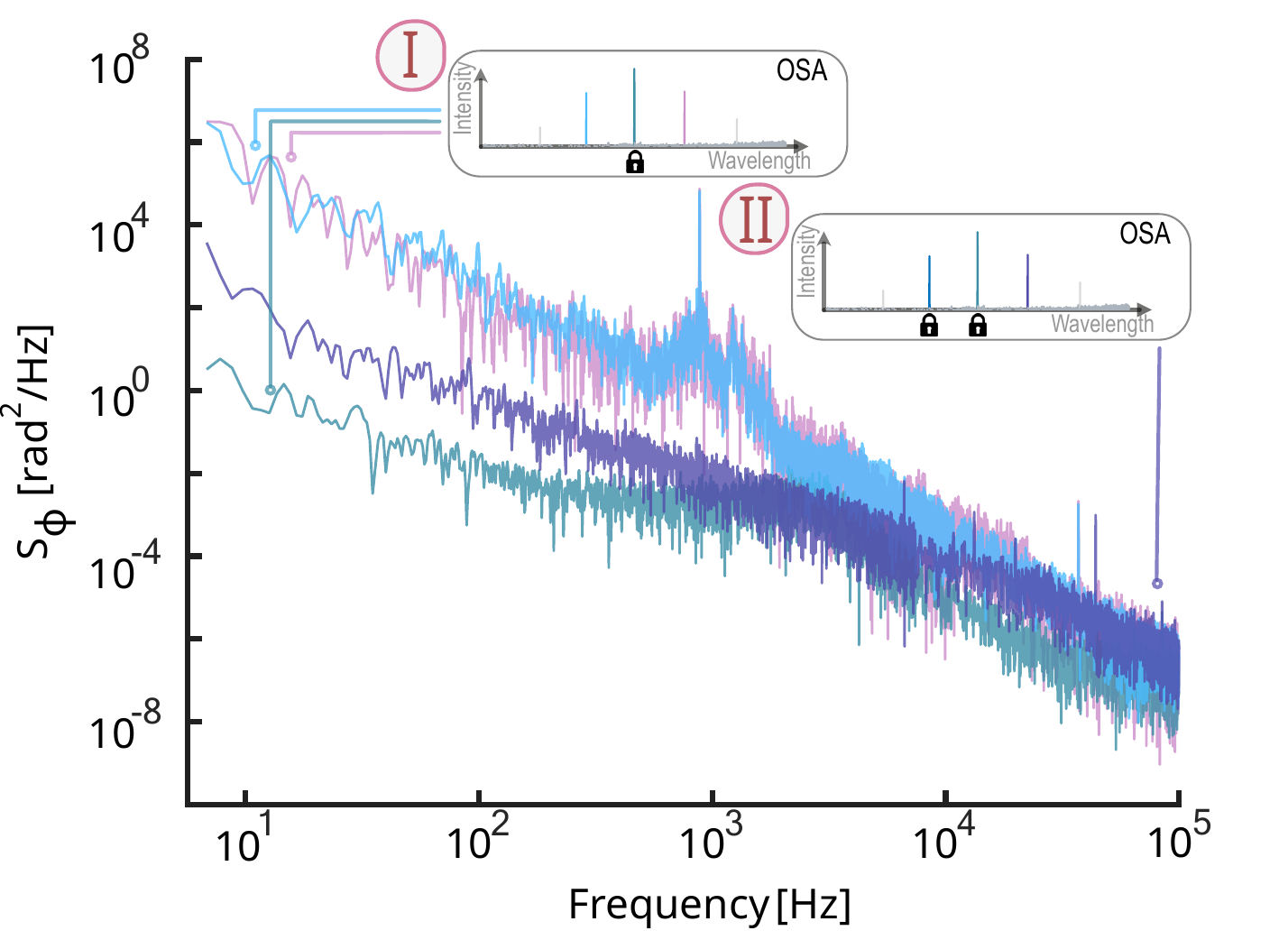}
    \caption{ \textbf{Phase-noise PSD $S_\phi(f)$ of the primary-comb
    heterodyne beats in the two locking regimes.} \textbf{Scenario~I} (weakly
    locked): servo-locked pump (light green) together with the corresponding signal
    (cyan) and idler (pink) tones, stabilized to the common DFC reference.
    \textbf{Scenario~II} (tightly locked): idler tone (navy blue) when both pump and
    signal are phase-locked. Insets: optical spectra (OSA) of the primary triad
    in the two regimes. Applying the $1~\mathrm{rad}^2$ integrated-linewidth
    criterion yields $\delta f\approx3$, $15$, and $17~\mathrm{kHz}$ for the
    pump, signal, and idler in Scenario~I, narrowing to $\sim9~\mathrm{kHz}$ for
    the idler in Scenario~II.}
    \label{fig5}
\end{figure}

Having characterized the long-term frequency stability through the Allan deviation, we now turn to the short-term coherence of the primary tones via their phase-noise spectra. Fig.~\ref{fig5} presents the phase-noise PSD $S_\phi(f)$ of the heterodyne beats in the two locking regimes. In Scenario~I (weakly locked), the pump (light green curve) is locked to the DFC via a servo loop while the signal is stabilized against the same reference using a delay-line architecture; the pump spectrum exhibits a distinct servo bump near its locking bandwidth, and the resulting signal and idler tones (cyan and pink curves) track each other closely across the entire spectral range. Scenario~II (tightly locked) focuses on the idler (navy blue curve): at low offset frequencies the tight lock suppresses the idler noise by approximately two orders of magnitude relative to Scenario~I, while above $\sim10^4$~Hz the two idler traces converge and share an identical trend.

We extract an integrated residual linewidth from the residual phase noise, defining the linewidth $\delta f$ as the lower integration limit for which the accumulated phase variance reaches one square radian \cite{hjelme1991semiconductor, lei2022optical}.
\begin{equation*}
    \int_{\delta f}^{\infty} S_\phi(f)\,df = 1~[\mathrm{rad}^2]
\end{equation*}
In the weakly locked regime this yields $\sim$3~kHz for the servo-locked pump and $\sim$15 and $\sim$17~kHz for the signal and idler, respectively. Tight locking narrows the idler to $\sim$9~kHz, nearly twofold coherence improvement that quantifies the benefit of active stabilization.

\section{Discussion and Summary}
We now discuss the technical and physical factors dictating the accuracy and limits of our synchronized measurement 
architecture. Although strong signal-idler correlations were robustly observed across multiple datasets, certain 
operating conditions exhibited a noticeable reduction in common-mode noise rejection. We note, that this reduced 
correlation occurred in regimes where the heterodyne beat notes maintained a relatively high signal-to-noise ratio 
(SNR) of approximately 35~dB within a 6~MHz band-pass filter. This can be attributed to both technical
tracking limitations and fundamental physical processes.

First, from a technical perspective, a high average SNR does not inherently guarantee error-free frequency tracking. If 
the micro-comb state becomes intrinsically noisy, such as when transitioning toward the chaotic modulation instability 
(MI) regime, instantaneous phase slips can occur. Under these conditions, the transient optical power fluctuations can 
cause the tracking oscillators or frequency counters to momentarily lose lock or miscount cycles, injecting 
uncorrelated technical noise into the measurement dataset despite a seemingly adequate time-averaged SNR.

Second, intrinsic physical mechanisms within the micro-resonator may decouple the signal and idler frequencies. A 
primary candidate is Raman scattering, which is inherently asymmetric between the Stokes-shifted idler and the anti-
Stokes signal and can therefore break the parametric symmetry of the pair \cite{karpov2016raman, lei2022optical}. A related effect, the Raman self-frequency 
shift, is well documented for dissipative Kerr solitons, where it redshifts the soliton spectral center and competes 
with the dispersive-wave-induced spectral recoil \cite{karpov2016raman}. Whether an analogous perline Raman asymmetry 
arises in the few-line primary regime which lacks a soliton's continuous envelope has not, to our knowledge, 
been established.

Beyond these localized noise sources, the present measurement framework also imposes practical limits on the interpretation of the residual fluctuations. In our synchronized heterodyne scheme, the measured beat-note records contain contributions from the local DFC lines, the RF detection electronics, the frequency counters, and the independent optical and fiber paths used for the signal and idler measurements. At short integration times, uncorrelated electronic noise in the photoreceivers and RF amplification chains can limit the resolution with which sub-Hz deviations in $\Delta$ are resolved. At longer timescales, uncompensated optical-path variations in the separate fiber runs to the signal and idler heterodyne setups can introduce slow differential drifts, which appear as imperfect correlation in the measured records.

To summarize, we have demonstrated a synchronized multi-channel measurement of the frequency correlations, relative stability, and residual coherence of primary-comb tones. First, under weakly locked, effectively free-running conditions with $\sim$500~kHz excursions, the energy-conservation residual remains sub-Hz in the mean, the regression slope deviates from the ideal $-1$ response by only $2.4\times10^{-9}$, and the signal-idler pair approaches near-perfect anti-correlation at long integration times, providing a direct measure of parametric rigidity. Second, this rigidity governs how stability is shared among the modes. By stabilizing two tones, maser-derived stability is transferred to the third, reaching roughly $6\times10^{-16}$ at $10^{2}$~s, while exposing a minute residual idler excess that is not fully suppressed by the common-mode constraint. The residual integrated linewidth further quantifies the coherence preserved by the parametric triad under tight locking. Taken together, these measurements show that primary Kerr tones behave as a frequency-rigid and mutually coherent parametric triad whose energy-conservation residual is preserved at a level suitable for demanding metrological and spectroscopic applications.

\bibliographystyle{unsrt} 
\bibliography{sample} 

@article{kippenberg2004kerr,
  author  = {Kippenberg, T. J. and Spillane, S. M. and Vahala, K. J.},
  title   = {Kerr-nonlinearity optical parametric oscillation in an ultrahigh-{Q} toroid microcavity},
  journal = {Phys. Rev. Lett.},
  volume  = {93},
  number  = {8},
  pages   = {083904},
  year    = {2004},
  doi     = {10.1103/PhysRevLett.93.083904}
}

@article{savchenkov2004low,
  author  = {Savchenkov, A. A. and Matsko, A. B. and Strekalov, D. and Mohageg, M. and Ilchenko, V. S. and Maleki, L.},
  title   = {Low threshold optical oscillations in a whispering gallery mode {CaF$_2$} resonator},
  journal = {Phys. Rev. Lett.},
  volume  = {93},
  number  = {24},
  pages   = {243905},
  year    = {2004},
  doi     = {10.1103/PhysRevLett.93.243905}
}

@article{delhaye2007optical,
  author  = {Del'Haye, P. and Schliesser, A. and Arcizet, O. and Wilken, T. and Holzwarth, R. and Kippenberg, T. J.},
  title   = {Optical frequency comb generation from a monolithic microresonator},
  journal = {Nature},
  volume  = {450},
  number  = {7173},
  pages   = {1214--1217},
  year    = {2007},
  doi     = {10.1038/nature06401}
}

@article{kippenberg2011microresonator,
  author  = {Kippenberg, T. J. and Holzwarth, R. and Diddams, S. A.},
  title   = {Microresonator-based optical frequency combs},
  journal = {Science},
  volume  = {332},
  number  = {6029},
  pages   = {555--559},
  year    = {2011},
  doi     = {10.1126/science.1193968}
}

@article{kippenberg2018dissipative,
  author  = {Kippenberg, T. J. and Gaeta, A. L. and Lipson, M. and Gorodetsky, M. L.},
  title   = {Dissipative {Kerr} solitons in optical microresonators},
  journal = {Science},
  volume  = {361},
  number  = {6402},
  pages   = {eaan8083},
  year    = {2018},
  doi     = {10.1126/science.aan8083}
}

@article{chembo2013spatiotemporal,
  author  = {Chembo, Y. K. and Menyuk, C. R.},
  title   = {Spatiotemporal {Lugiato-Lefever} formalism for {Kerr}-comb generation in whispering-gallery-mode resonators},
  journal = {Phys. Rev. A},
  volume  = {87},
  number  = {5},
  pages   = {053852},
  year    = {2013},
  doi     = {10.1103/PhysRevA.87.053852}
}

@article{herr2012universal,
  author  = {Herr, T. and Hartinger, K. and Riemensberger, J. and Wang, C. Y. and Gavartin, E. and Holzwarth, R. and Gorodetsky, M. L. and Kippenberg, T. J.},
  title   = {Universal formation dynamics and noise of {Kerr}-frequency combs in microresonators},
  journal = {Nat. Photonics},
  volume  = {6},
  number  = {7},
  pages   = {480--487},
  year    = {2012},
  doi     = {10.1038/nphoton.2012.127}
}

@article{godey2014stability,
  author  = {Godey, C. and Balakireva, I. V. and Coillet, A. and Chembo, Y. K.},
  title   = {Stability analysis of the spatiotemporal {Lugiato-Lefever} model for {Kerr} optical frequency combs in the anomalous and normal dispersion regimes},
  journal = {Phys. Rev. A},
  volume  = {89},
  number  = {6},
  pages   = {063814},
  year    = {2014},
  doi     = {10.1103/PhysRevA.89.063814}
}

@article{gaeta2019photonic,
  author  = {Gaeta, A. L. and Lipson, M. and Kippenberg, T. J.},
  title   = {Photonic-chip-based frequency combs},
  journal = {Nat. Photonics},
  volume  = {13},
  number  = {3},
  pages   = {158--169},
  year    = {2019},
  doi     = {10.1038/s41566-019-0358-x}
}

@article{pasquazi2018microcombs,
  author  = {Pasquazi, A. and Peccianti, M. and Razzari, L. and Moss, D. J. and Coen, S. and Erkintalo, M. and Chembo, Y. K. and Hansson, T. and Wabnitz, S. and Del'Haye, P. and Xue, X. and Weiner, A. M. and Morandotti, R.},
  title   = {Micro-combs: a novel generation of optical sources},
  journal = {Phys. Rep.},
  volume  = {729},
  pages   = {1--81},
  year    = {2018},
  doi     = {10.1016/j.physrep.2018.01.004}
}

@article{liang2015high,
  author  = {Liang, W. and Eliyahu, D. and Ilchenko, V. S. and Savchenkov, A. A. and Matsko, A. B. and Seidel, D. and Maleki, L.},
  title   = {High spectral purity {Kerr} frequency comb radio frequency photonic oscillator},
  journal = {Nat. Commun.},
  volume  = {6},
  pages   = {7957},
  year    = {2015},
  doi     = {10.1038/ncomms8957}
}

@article{huang2015low,
  author  = {Huang, Shu-Wei and Yang, Jinghui and Lim, Jinkang and Zhou, Heng and Yu, Mingbin and Kwong, Dim-Lee and Wong, Chee Wei},
  title   = {A low-phase-noise 18 {GHz} {Kerr} frequency microcomb phase-locked over 65 {THz}},
  journal = {Sci. Rep.},
  volume  = {5},
  pages   = {13355},
  year    = {2015},
  doi     = {10.1038/srep13355}
}

@article{li2012low,
  author  = {Li, J. and Lee, H. and Chen, T. and Vahala, K. J.},
  title   = {Low-pump-power, low-phase-noise, and microwave to millimeter-wave repetition rate operation in microcombs},
  journal = {Phys. Rev. Lett.},
  volume  = {109},
  number  = {23},
  pages   = {233901},
  year    = {2012},
  doi     = {10.1103/PhysRevLett.109.233901}
}

@article{pfeifle2015optimally,
  author  = {Pfeifle, J. and Coillet, A. and Henriet, R. and Saleh, K. and Schindler, P. and Weimann, C. and Freude, W. and Balakireva, I. V. and Larger, L. and Koos, C. and Chembo, Y. K.},
  title   = {Optimally coherent {Kerr} combs generated with crystalline whispering gallery mode resonators for ultrahigh capacity fiber communications},
  journal = {Phys. Rev. Lett.},
  volume  = {114},
  number  = {9},
  pages   = {093902},
  year    = {2015},
  doi     = {10.1103/PhysRevLett.114.093902}
}

@article{coillet2014robustness,
  author  = {Coillet, A. and Chembo, Y. K.},
  title   = {On the robustness of phase locking in {Kerr} optical frequency combs},
  journal = {Opt. Lett.},
  volume  = {39},
  number  = {6},
  pages   = {1529--1532},
  year    = {2014},
  doi     = {10.1364/OL.39.001529}
}

@article{huang2019thermorefractive,
  author  = {Huang, G. and Lucas, E. and Liu, J. and Raja, A. S. and Lihachev, G. and Gorodetsky, M. L. and Engelsen, N. J. and Kippenberg, T. J.},
  title   = {Thermorefractive noise in silicon-nitride microresonators},
  journal = {Phys. Rev. A},
  volume  = {99},
  number  = {6},
  pages   = {061801},
  year    = {2019},
  doi     = {10.1103/PhysRevA.99.061801}
}

@article{kondratiev2018thermorefractive,
  author  = {Kondratiev, N. M. and Gorodetsky, M. L.},
  title   = {Thermorefractive noise in whispering gallery mode microresonators: Analytical results and numerical simulation},
  journal = {Phys. Lett. A},
  volume  = {382},
  number  = {33},
  pages   = {2265--2268},
  year    = {2018},
  doi     = {10.1016/j.physleta.2017.04.043}
}

@article{schunemann1999simple,
  author  = {Sch\"unemann, U. and Engler, H. and Grimm, R. and Weidem\"uller, M. and Zielonkowski, M.},
  title   = {Simple scheme for tunable frequency offset locking of two lasers},
  journal = {Rev. Sci. Instrum.},
  volume  = {70},
  number  = {1},
  pages   = {242--243},
  year    = {1999},
  doi     = {10.1063/1.1149573}
}

@article{dawkins2007considerations,
  author  = {Dawkins, S. T. and McFerran, J. J. and Luiten, A. N.},
  title   = {Considerations on the measurement of the stability of oscillators with frequency counters},
  journal = {IEEE Trans. Ultrason. Ferroelectr. Freq. Control},
  volume  = {54},
  number  = {5},
  pages   = {918--925},
  year    = {2007},
  doi     = {10.1109/TUFFC.2007.337}
}

@inproceedings{groslambert1981characterization,
  author    = {Groslambert, J. and Fest, D. and Olivier, M. and Gagnepain, J. J.},
  title     = {Characterization of frequency fluctuations by crosscorrelations and by using three or more oscillators},
  booktitle = {Proc. 35th Annual Frequency Control Symposium},
  pages     = {458--463},
  year      = {1981},
  doi       = {10.1109/FREQ.1981.200513}
}

@article{vernotte2019three,
  author  = {Vernotte, F. and Lantz, E.},
  title   = {Three-cornered hat and {Groslambert} covariance: A first attempt to assess the uncertainty domains},
  journal = {IEEE Trans. Ultrason. Ferroelectr. Freq. Control},
  volume  = {66},
  number  = {3},
  pages   = {643--653},
  year    = {2019},
  doi     = {10.1109/TUFFC.2018.2884370}
}

@article{hjelme1991semiconductor,
  author  = {Hjelme, D. R. and Mickelson, A. R. and Beausoleil, R. G.},
  title   = {Semiconductor laser stabilization by external optical feedback},
  journal = {IEEE J. Quantum Electron.},
  volume  = {27},
  number  = {3},
  pages   = {352--372},
  year    = {1991},
  doi     = {10.1109/3.81333}
}

@article{lei2022optical,
  author  = {Lei, Fuchuan and Ye, Zhichao and Helgason, {\'O}skar B. and F{\"u}l{\"o}p, Attila and Girardi, Marcello and Torres-Company, Victor},
  title   = {Optical linewidth of soliton microcombs},
  journal = {Nat. Commun.},
  volume  = {13},
  number  = {1},
  pages   = {3161},
  year    = {2022},
  doi     = {10.1038/s41467-022-30726-5}
}

@article{karpov2016raman,
  author  = {Karpov, Maxim and Guo, Hairun and Kordts, Arne and Brasch, Victor and Pfeiffer, Martin H. P. and Zervas, Michalis and Geiselmann, Michael and Kippenberg, Tobias J.},
  title   = {Raman self-frequency shift of dissipative {Kerr} solitons in an optical microresonator},
  journal = {Nat. Phys.},
  volume  = {12},
  number  = {1},
  pages   = {53--58},
  year    = {2016},
  doi     = {10.1038/nphys3503}
}

@article{Diallo2017PrimaryKerrCombs,
  author  = {Diallo, Souleymane and Chembo, Yanne K.},
  title   = {Optimization of primary {Kerr} optical frequency combs for tunable microwave generation},
  journal = {Optics Letters},
  volume  = {42},
  number  = {18},
  pages   = {3522--3525},
  year    = {2017},
  month   = sep,
  doi     = {10.1364/OL.42.003522},
  url     = {https://doi.org/10.1364/OL.42.003522}
}

\end{document}